A Chromatographic Process Design and Optimization Platform Powered by Large Language Models: A Case Application on Extract of *Ginkgo Biloba* Leaf

Zhilong Tang[1], Shaohua Wu[1], Xinyan Zhao[2], Yu Wang[1,3], Xingchu Gong[1,2,3,4,*]

1. Pharmaceutical Informatics Institute, College of Pharmaceutical Sciences, Zhejiang University, Hangzhou 310058, China

2. Institute of Traditional Chinese Medicine, Tianjin University of Traditional Chinese Medicine, Tianjin, 301617, China

3. Jinhua Institute of Zhejiang University, Jinhua 321016, China

4. National Key Laboratory of Chinese Medicine Modernization, Zhejiang University, Hangzhou 310058, China

[*] Correspondence: gongxingchu@zju.edu.cn; Tel.: +86-571-88208427

Abstract: Chromatographic separation technology has been widely applied in pharmaceutical, chemical, and food industries due to its high efficiency. However, traditional human-dependent chromatographic process development faces challenges such as reliance on expert experience, long development cycles, and labor intensity. ChromR, a large language model (LLM)-driven platform for chromatographic process design and optimization, is presented in this work. The platform integrates ChromLLM, a domain-specific LLM trained for chromatography, along with a multi-agent system and an automated chromatographic experimental device. The multi-agent system comprises four agents: domain knowledge answering, experimental design, experimental execution, and data analysis. ChromR enables automatic completion of the entire workflow—including initial process parameter recommendation, experimental design, automated execution, data analysis, and multi-objective optimization. By utilizing ChromR, dependency on expert knowledge is effectively reduced, while labor input and development time are significantly decreased. Chromatographic purification of the extract of *Ginkgo biloba* leaf (EGBL) was selected as a case study. ChromR successfully developed a chromatographic process within one week that meets multiple objectives, including fraction quality and production efficiency, reducing development time to approximately 1/7 of that required by the conventional paradigm. An intelligent, automated, and universally applicable new paradigm was established for chromatographic process development.

1. Introduction

Chromatographic separation technology, owing to its advantages of high efficiency, has been widely employed in separation and purification processes across chemical, pharmaceutical, and food industries[1-5]. Chromatographic processes are complex, and the separation performance of target compounds is generally determined by multiple parameters with strong inter-parameter coupling, making parameter optimization challenging[6, 7]. Achieving satisfactory separation alone is insufficient in chromatographic process development. Considerations must also include cost[8, 9], green chemistry principles[10, 11], and efficiency[12]. These multi-objective requirements result in more challenges in chromatographic process development.

Traditional chromatographic process development typically involves literature retrieval, stationary phase screening[13, 14], one-factor-at-a-time (OFAT) investigation[15, 16], design of experiment (DOE) investigation[17, 18], modelling-based optimization, and parameter validation[19, 20]. Even for experienced researchers, developing a satisfactory chromatographic process remains time-consuming and labor-intensive. Therefore, it is necessary to develop a more efficient approach for chromatographic process development.

In recent years, self-driving laboratories (SDLs) have emerged in research fields such as chemistry[21-25], biology[26-28], and materials science[29], and have proven to be effective tools for accelerating scientific discovery[30]. SDLs generate high-quality, high-density experimental data through automated experimentation. The data are transmitted to an AI decision-making system for analysis, which updates predictive models and suggests the most promising next experimental steps, forming a closed-loop iteration until predefined research goals are achieved[31]. Bayesian optimization (BO) is frequently employed as the experimental planning algorithm within the AI decision-making system[31, 32]. However, BO is sensitive to initial conditions[33], and existing non-informative sampling methods (e.g., random or Latin hypercube sampling[34]) often neglect prior knowledge available in the literature, leading to wastage of resources.

Since the release of large language models (LLMs) represented by ChatGPT, their outstanding performance in natural language processing has prompted scientists to explore their application potential in research fields such as chemistry, biology, and materials science[35], for example, literature data mining[36-38], chemical and material discovery[39, 40], protein and gene sequence design[26, 41], and laboratory automation[42, 43]. After training on vast corpora, LLMs can learn complex linguistic rules, semantics, and contextual associations within the data, enabling them to perform reasonable inference based on prior knowledge upon receiving complex or non-standardized inputs from researchers and return appropriate responses[44]. Many researchers across various domains have already employed domain-specific corpora to train LLMs specialized in their respective fields[45-47].

Currently, LLMs still struggle with performing complex mathematical computations. However, this limitation can be addressed by developing agents that integrate multiple tools with an LLM serving as the central coordinator[48, 49]. Through tool-augmented LLMs, it becomes possible to directly operate hardware in the physical world via predefined API interfaces according to needs, thereby significantly reducing human labor and accelerating scientific discovery[50].

A chromatographic process design and optimization platform driven by an LLM, named ChromR, was developed to assist researchers in accelerating chromatographic process development in this work. Extract of *Ginkgo biloba* leaf (EGBL) exhibits various pharmacological activities, including cognitive function improvement[51], antioxidant effects[52], and anti-inflammatory properties[53], among others. Therefore, EGBL was selected as a case example to demonstrate the workflow of ChromR and its performance in chromatographic process design and optimization.

2. Methods

2.1 Overview of ChromR

ChromR consists of two components: a multi-agent system and an automated chromatography experimental device. The multi-agent system includes four agents responsible for chromatography domain knowledge question-answering, DOE, experimental execution, and data analysis, respectively. In this work, we trained for the first time a large language model specialized in the chromatography domain (ChromLLM), based on which an agent (Agent A) was developed for initial chromatographic process design. The detailed architecture of ChromR is illustrated in Fig. 1.

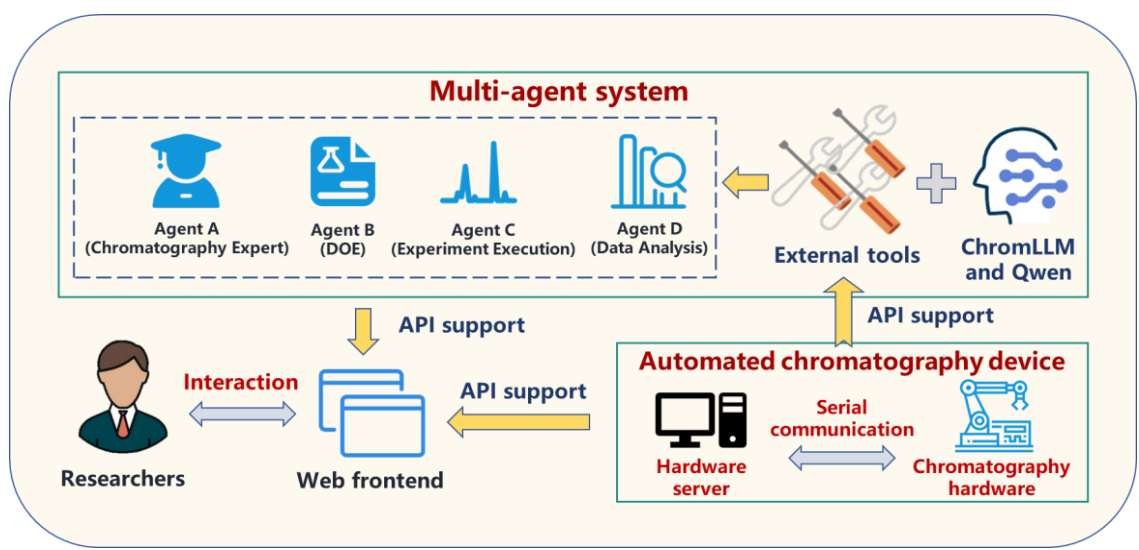

**Fig. 1.** Schematic illustration of ChromR.

2.2 Construction of ChromLLM

2.2.1 Data collection

In traditional data mining workflows, regular expressions are powerful and commonly used tools. However, this approach is labor-intensive and requires researchers to possess substantial expertise in programming, computer science, and data analysis[54, 55]. In this work, to maximize information extraction from the literature while minimizing manual effort, we established LLM-based pipelines specifically designed for literature data mining.

The data used for post-pretraining belong to unannotated plain text. The specific pipeline for unlabeled data mining is shown in Fig. 2A. We first retrieved and downloaded a large number of chromatography-related literature from databases, then performed data mining using the Qwen-Long model, which supports up to 10 million tokens. Batch inference was conducted on the Alibaba Cloud Bailian platform (https://bailian.console.aliyun.com). Before batch inference, prompt engineering was applied to optimize the response of Qwen-Long, ensuring the quality and accuracy of the extracted data. The inference results often contained invisible symbols, personal information, garbled characters, and lengthy repetitions. Thus, we employed data cleaning techniques such as regular expressions and similarity-based deduplication to ensure the quality of training data.

The specific pipeline for acquiring annotated data for fine-tuning is shown in Fig. 2B. We first selected several representative literature and, after optimizing prompt engineering, used the Qwen-Long model to perform batch inference to obtain supervised fine-tuning data. After splitting the data into question-answer pairs, similar data cleaning procedures were applied. Notably, after data cleaning, human experts further reviewed and revised the data to ensure high quality and alignment with human preferences.

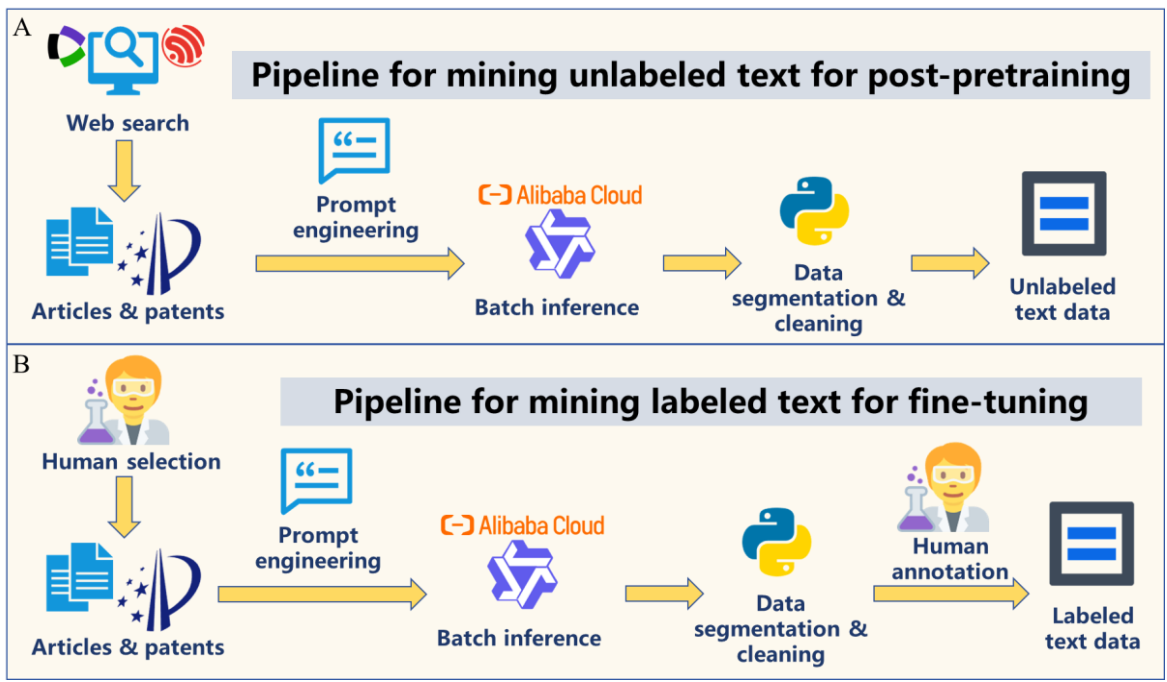

**Fig. 2.** Pipelines for data mining. (A) Collection of unlabeled data for post-pretraining. (B) Collection of labelled data for supervised fine-tuning.

2.2.2 Post-pretraining

We obtained a dataset totalling over 75 million tokens using the pipeline described in Section 2.1.1. Qwen2.5-14B was selected as the base model. Model training was carried out on the Huawei Cloud (https://console.huaweicloud.com/modelarts). The learning rate was set to 0.00002, and the number of epochs was set to 1. The training loss curve is shown in Fig. S1. Since the loss had converged close to zero, we concluded that the model had sufficiently learned the knowledge embedded in the data.

2.2.3 Supervised fine-tuning

After manual inspection, we obtained a total of 6,801 entries suitable for fine-tuning, most of which belonged to multi-turn dialogues. This work performed Low-Rank Adaptation (LoRA) fine-tuning on the post-pretrained LLM using Llama-Factory 0.9.2.dev0 on a computer equipped with four NVIDIA GeForce RTX 3090 GPUs. The learning rate was set to 0.00005, and the number of epochs was set to 1. The loss curve during supervised fine-tuning is shown in Fig. S2. The loss converged to approximately 0.8, indicating that the model had learned most of the knowledge present in the data.

2.3 Multi-agent system

The multi-agent system comprises four agents with functions including chromatography domain knowledge answering, DOE, experimental execution, and data analysis. The system can receive prompt input from researchers and uploaded files, and intelligently determine which agents need to be activated based on the researchers' questions or requirements. Activated agents then intelligently invoke integrated tools according to the input prompts and generate responses. The system employs Qwen-series LLMs[56]for intent recognition and planning through prompt engineering. The multi-agent system was deployed locally on the Dify 1.5.1 platform.

2.3.1 Agent A: domain knowledge answering

Agent A consists of three main components: ChromLLM, a knowledge base, and a literature retrieval module. ChromLLM is capable of answering the majority of professional inquiries. To enhance response accuracy, a local knowledge base encompassing multiple standards and regulations to support retrieval-augmented generation (RAG) was established in this work. The literature retrieval module, driven by LLMs, can connect to academic databases including arXiv, PubMed, and Scopus. It can generate search queries based on researchers' inputs, retrieve relevant literature, and return the most pertinent data, thereby improving the advancement and timeliness of responses. After receiving outputs from the three components, a high-performance LLM from the Qwen series synthesizes the information and delivers the final answer to the researcher.

2.3.2 Agent B: DOE

Agent B intelligently identifies critical information from user prompts and invokes a Python executor to generate experimental designs. Agent B can generate common experimental designs, including Definitive Screening Design (DSD), Box-Behnken Design, Central Composite Design and so on.

2.3.3 Agent C: experimental execution

Agent C interprets researchers' natural language descriptions, generates hardware control code, and executes individual or sequential experiments (e.g., an entire experimental design). Embedded with API interfaces that directly operate physical hardware, Agent C automatically generates control commands based on analysis of researchers' requirements and transmits them to the hardware, accelerating the realization of research ideas in experimental practice.

2.3.4 Agent D: data analysis

Agent D integrates multiple data analysis tools, including stepwise regression, design space computation[57], and multi-objective optimization based on NSGA-II/III algorithms. During stepwise regression, it not only generates polynomial models but also analyzes the models and provides improvement recommendations.

2.4 Automated chromatography experimental device

Although numerous commercial automated chromatography instruments are available, their control protocols and operational methods are largely proprietary, making it difficult to achieve comprehensive bidirectional communication between all chromatographic units and computer controllers[31, 58]. To integrate with the multi-agent system and enable LLM-driven design and optimization of chromatographic processes, a custom automated chromatography experimental device was developed, as shown in Fig. 3 (photograph in Fig. S3). The entire system is controlled by a hardware server via serial communication. Two peristaltic pumps, P1–P2 (DIPump550-B146, Shanghai Kamoer Fluid Tech Co., Ltd), are installed before and after the chromatography column, respectively. Valves, V1–V9 (DN8 ball valves, Jian Kaiqiang Valve Trading Department), are computer-controlled to enable flexible switching between different chromatographic process stages. The device incorporates three types of online sensors: a pH meter (LN-ISEP20L05), a conductivity meter (LN-ISEP10L05), and an ORP meter (LN-ISRC20L05), all purchased from Shanghai Lanchang Auto Technology Co., Ltd.; and two sets of online spectrometers from Shenzhen South-Instrument Technology Co., Ltd., including a UV spectrometer (UV, SPEC-CDS350) and a near-infrared spectrometer (NIR, NIR-F210). Sensor and spectrometer data are stored in a MySQL database and used to determine the completion of regeneration and equilibration stages. Additionally, to maintain stable liquid levels within the chromatography column, an infrared level sensor (TOF-2m, Chengdu Zhongshan Technology Co., Ltd.) was installed above the column. The flow rate of pump P2 was regulated using a fuzzy controller according to the measured liquid level. The source code for system control is available at https://github.com/tzl1125/ChromR.

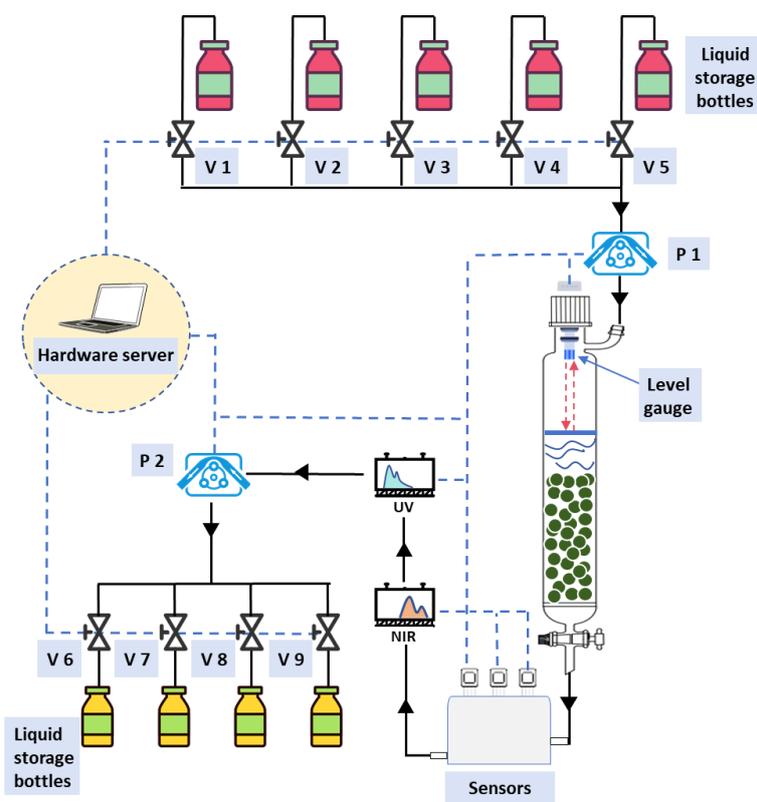

**Fig. 3.** Automated chromatography experimental device (solid lines represent fluid pathways, and dashed lines represent serial communication).

2.5 Interaction with ChromR

To facilitate the use of ChromR, a user-friendly web application was also developed in this work, as shown in Fig. S4. The front-end graphical interface was developed using Vue3.js and Node.js frameworks. Through the front-end interface, researchers can not only interact with agents to execute experiments but also directly manipulate each unit of the device. On the back-end, the FastAPI framework was employed to manage diverse tasks of the multi-agent system and experimental device, including tools' API calls, sensor data management, and transmission of control instructions. All source code is available at https://github.com/tzl1125/ChromR.

2.6 Applications of ChromR

ChromR was tested using the chromatographic purification process of EGBL as an example. In the EGBL purification process, macroporous resin serves as the separation medium. The process involves feeding with aqueous precipitation solution, washing with low-concentration ethanol solution, and elution of the final fraction solution with high-concentration ethanol solution. According to the 2025 edition of Volume I of the Chinese Pharmacopoeia, after column chromatography treatment, the content of flavonoid glycosides (FG) in EGBL must be no less than 24.0%, and terpene trilactones (TT) no less than 6.0%[59].

The process objectives defined during optimization include: maximizing the purities and productivities of FG and TT while meeting pharmacopoeia standards. The formulas for calculating FG and TT purities are given in Equation (1), and those for FG and TT productivities are given in Equation (2):

$$\text{Purity} = \frac{m_{target}}{m_{ts}} \times 100\% \qquad (1)$$

$$\text{Productivity} = \frac{m_{tt}}{t} \qquad (2)$$

where $m_{target}$ is the mass of the target substance per unit volume of fraction (mg), $m_{ts}$ is the total solid mass per unit volume of fraction (mg), $m_{tt}$ is the total mass of the target substance in the fraction obtained from one chromatographic purification cycle (mg), and $t$ is the time consumed during one chromatographic purification process (h). The duration of the equilibration and regeneration steps was not considered during optimization. We investigated the effects of six process parameters on these objectives: feed flow, feed volume, wash flow, wash volume, elution flow, and elution volume. The equilibration and regeneration flows were fixed at 3 BV/h, while the equilibration and regeneration volumes were determined based on sensor and spectral signals. Once the signals stabilized, the system automatically transitioned to the next phase. Water was used as the equilibration solvent, and 95% (v/v) ethanol was used as the regeneration solvent.

The main steps involved in interacting with ChromR include: (1) recommending initial values for the process parameters under investigation, along with other necessary parameters such as column bed aspect ratio, resin type, wash solvent, and elution solvent; (2) defining the ranges of the process parameters to be studied for experimental design; (3) automated execution of experiments according to the experimental design table; (4) analysis of experimental data, model development, and process parameter optimization; and (5) validation of the optimal process.

To better evaluate the performance of ChromR, 13 batches of *Ginkgo biloba* ethanolic extract followed by water precipitation were prepared as feed solutions. Four parameters—concentrations and contents of FG and TT—were selected as material attributes. The material attributes of each batch are listed in Table S1. Fig. 4 presents the material attributes and variations among the 13 batches of feed solutions. Ten of these batches were used for process optimization, while the remaining three batches—231201, 250409, and 250401—were reserved for validating the optimized process. Detailed experimental methods are provided in Section S5.

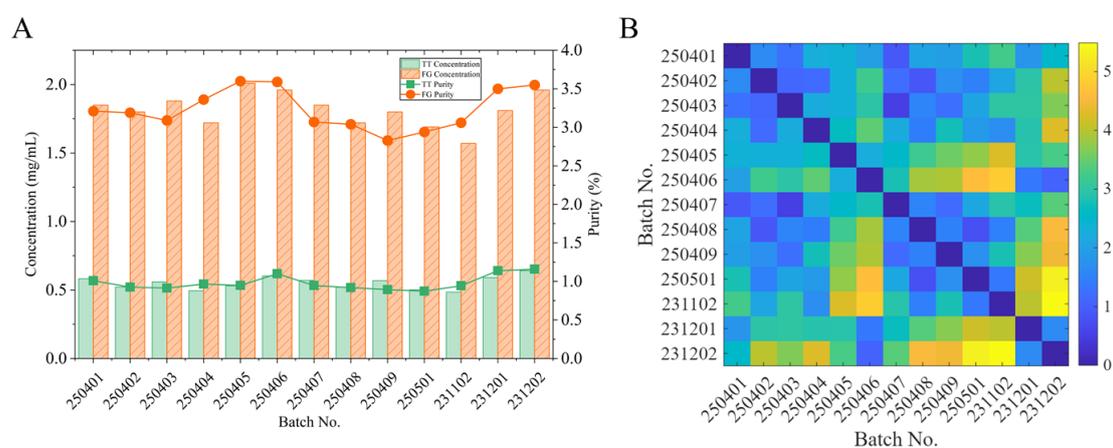

**Fig. 4.** Attributes and differences of feed solutions across batches. (A) Material attributes of 13 batches of feed solutions. (B) Euclidean distance matrix between feed solution batches.

## 3. Results

### 3.1 Initial chromatographic process conditions

A gradual approach was adopted to allow ChromR to recommend initial process conditions: first defining the resin type, wash solvent, and elution solvent, followed by other process parameters.

Through prompt engineering, we informed ChromR of the constraints and optimization objectives of the process, enabling it to recommend suitable resin types and ethanol concentrations for the wash and elution solvents. The detailed interaction process is shown in Figs. S6 and S7.

ChromR successfully retrieved two relevant literature[60, 61] and regulations from the Chinese Pharmacopoeia. ChromR first recommended AB-8 macroporous adsorption resin, providing justification based on chemical property analysis between the resin and the target compounds. ChromR also noted that XDA-1 resin was preferred in the literature; however, due to the advantages of AB-8 in cost and stability, ChromR ultimately selected AB-8 resin. This decision aligns with reported findings[62, 63], although ChromR did not retrieve these specific references, demonstrating the strong reasoning capability of the built-in ChromLLM. Subsequently, based on the principle of "like dissolves like", ChromR recommended 20% (v/v) aqueous ethanol as the wash solvent and 75% (v/v) aqueous ethanol as the elution solvent.

During the training phase of the LLM, a large number of chromatographic process conditions containing numerical parameters were specifically collected from the literature to enhance ChromLLM's ability to accurately recommend numerical process parameters. Similarly, via prompting, we provided ChromR with the already-determined process parameters, feed solution concentration, and process objectives, and requested recommendations for other specific process parameters, including column bed aspect ratio and flow rates and durations for each chromatographic phase. The detailed Q&A process is shown in Figs. S8–S10.

Although ChromR did not retrieve any literature, with assistance from ChromLLM and the knowledge base, it still provided reasonable recommendations. First, considering column packing stability, flow uniformity, and elution resolution, ChromR proposed a bed aspect ratio of 12:1. Then, based on the process objectives of purity and productivity for FG and TT, ChromR recommended the following parameters: feed time of 1.5 h, wash time of 1 h, elution time of 1 h, feed flow of 1 BV/h, wash flow of 2 BV/h, and elution flow of 3 BV/h.

3.2 Experimental design and execution

3.2.1 DOE

After defining the initial chromatographic process parameters, we instructed ChromR to generate a DSD experimental design table to further investigate response surface models for each performance indicator. We set the number of virtual factors to 2 and added 3 center point experiments. Ten batches of feed solutions were selected for process optimization in this work. ChromR employed a uniform and random allocation method to assign batch numbers to the experimental design table. The detailed interaction process is shown in Fig. S11, and the DSD experimental design table generated by ChromR is presented in Table 1.

**Table 1.** Experimental design table and results.

| No. | Feed flow $X_1$ (BV/h) | Feed time $X_2$ (h) | Wash flow $X_3$ (BV/h) | Washing time $X_4$ (h) | Elution flow $X_5$ (BV/h) | Elution time $X_6$ (h) | Feed solution batch No. | TT purity $Y_1$ (%) | TT productivity $Y_2$ (mg/h) | FG purity $Y_3$ (%) | FG productivity $Y_4$ (mg/h) |
|---|---|---|---|---|---|---|---|---|---|---|---|
| 1 | 1.0 | 2.0 | 2.5 | 1.5 | 3.5 | 1.5 | 250408 | 7.18 | 39.3 | 45.0 | 247 |
| 2 | 1.0 | 1.0 | 1.5 | 0.5 | 2.5 | 0.5 | 250403 | 1.14 | 31.5 | 6.65 | 184 |
| 3 | 1.5 | 1.5 | 1.5 | 0.5 | 3.5 | 0.5 | 231102 | 2.80 | 111 | 9.71 | 384 |
| 4 | 0.5 | 1.5 | 2.5 | 1.5 | 2.5 | 1.5 | 231202 | 7.58 | 23.3 | 30.3 | 93.0 |
| 5 | 1.5 | 2.0 | 2.0 | 0.5 | 2.5 | 1.5 | 250405 | 4.46 | 102 | 17.1 | 391 |
| 6 | 0.5 | 1.0 | 2.0 | 1.5 | 3.5 | 0.5 | 250405 | 6.87 | 16.0 | 31.5 | 73.1 |
| 7 | 1.5 | 2.0 | 2.5 | 1.0 | 2.5 | 0.5 | 250404 | 6.46 | 57.4 | 27.8 | 247 |
| 8 | 0.5 | 1.0 | 1.5 | 1.0 | 3.5 | 1.5 | 250402 | 4.47 | 16.4 | 21.6 | 79.2 |
| 9 | 1.5 | 1.0 | 2.5 | 1.5 | 3.0 | 0.5 | 250408 | 8.70 | 51.1 | 36.9 | 217 |
| 10 | 0.5 | 2.0 | 1.5 | 0.5 | 3.0 | 1.5 | 250406 | 1.79 | 29.7 | 8.14 | 134 |
| 11 | 1.5 | 2.0 | 1.5 | 1.5 | 3.5 | 1.0 | 250402 | 8.16 | 90.7 | 32.2 | 358 |

| 12 | 0.5 | 1.0 | 2.5 | 0.5 | 2.5 | 1.0 | 250407 | 3.62 | 23.4 | 14.4 | 93.4 |
| 13 | 1.5 | 1.0 | 2.5 | 0.5 | 3.5 | 1.5 | 250406 | 4.59 | 59.9 | 20.8 | 271 |
| 14 | 0.5 | 2.0 | 1.5 | 1.5 | 2.5 | 0.5 | 231202 | 4.62 | 18.8 | 21.8 | 89.0 |
| 15 | 1.5 | 1.0 | 1.5 | 1.5 | 2.5 | 1.5 | 250501 | 7.31 | 49.9 | 29.7 | 203 |
| 16 | 0.5 | 2.0 | 2.5 | 0.5 | 3.5 | 0.5 | 250403 | 4.20 | 38.4 | 17.8 | 162 |
| 17 | 1.0 | 1.5 | 2.0 | 1.0 | 3.0 | 1.0 | 250407 | 6.48 | 56.7 | 25.8 | 226 |
| 18 | 1.0 | 1.5 | 2.0 | 1.0 | 3.0 | 1.0 | 250501 | 7.06 | 57.4 | 28.4 | 231 |
| 19 | 1.0 | 1.5 | 2.0 | 1.0 | 3.0 | 1.0 | 250404 | 7.41 | 54.4 | 29.2 | 214 |
| 20 | 1.0 | 1.5 | 2.0 | 1.0 | 3.0 | 1.0 | 231102 | 6.97 | 65.9 | 24.3 | 230 |

### 3.2.2 Experiment execution

For the initial process conditions recommended by ChromR, we first directed ChromR to automatically conduct four experiments with identical operating conditions (experiments 17, 18, 19, and 20). The results are shown in Table 1. Despite significant compositional differences among the four feed solution batches, the fractions obtained under the initial process conditions recommended by ChromR already met the purity requirements for TT and FG specified in the Chinese Pharmacopoeia. The productivity of TT exceeded 50 mg/h, and that of FG exceeded 200 mg/h in all cases.

Subsequently, we instructed ChromR to automatically execute all remaining experiments in the experimental design table. The dialogue interface is shown in Figs. S12–S14. Due to automation, ChromR operated unattended overnight, completing all experiments within less than one week, significantly reducing labor and time costs.

### 3.3 Process optimization

### 3.3.1 Model development

After the experiments were completed, we instructed ChromR to build models using stepwise regression (p=0.05). The specific dialogue content is shown in Figs. S15 and S16. ChromR successfully established models for each performance indicator based on data obtained from automated experimental runs, with contour plots presented in Fig. 5. The partial regression coefficients of each model are listed in Table S2, and the fitting results are shown in Fig. S17. As illustrated in Fig. 5, increasing wash time and wash flow contribute to higher purities of TT and FG, whereas increasing feed time and feed flow enhance the productivity of TT and FG.

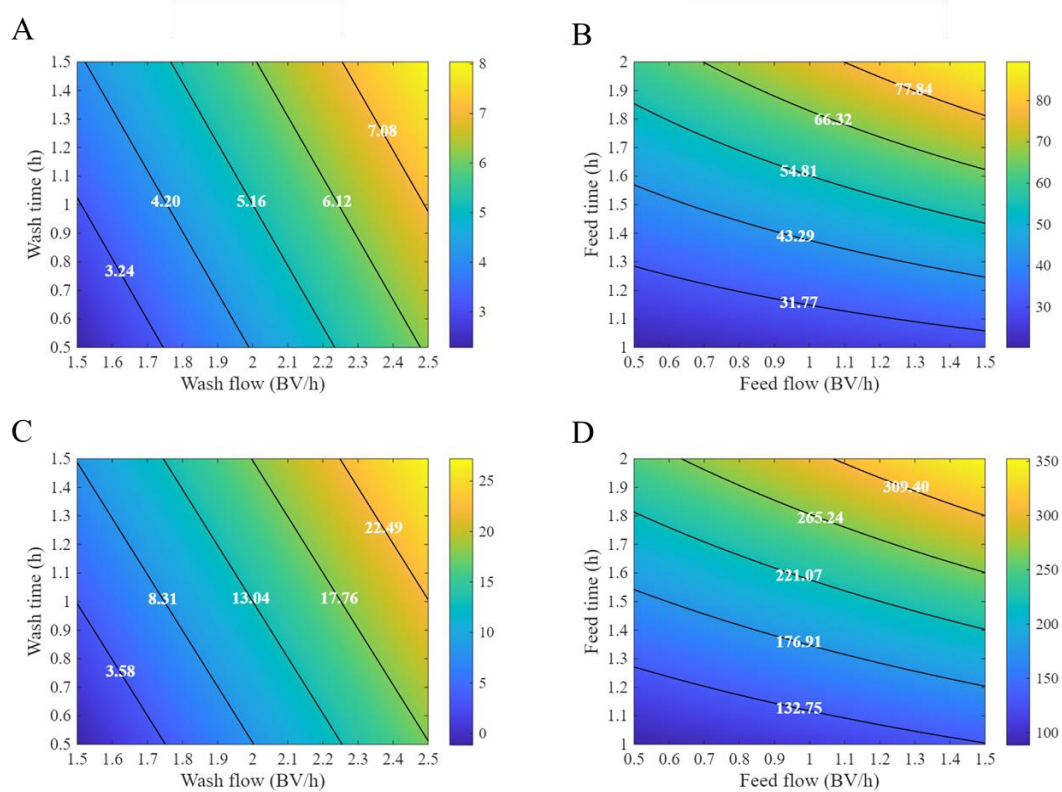

**Fig. 5**. Contour plots for Batch 250401. (A) TT purity, $X_1$ = 1.0 BV/h, $X_5$ = 3.0 BV/h, $X_6$ = 1.0 h. (B) TT productivity, $X_3$ = 2.0 BV/h, $X_4$ = 1.0 h, $X_5$ = 3.0 BV/h, $X_6$ = 1.0 h. (C) FG purity, $X_1$ = 1.0 BV/h, $X_5$ = 3.0 BV/h, $X_6$ = 1.0 h. (D) FG productivity, $X_3$ = 2.0 BV/h, $X_4$ = 1.0 h, $X_5$ = 3.0 BV/h, $X_6$ = 1.0 h.

The $R^2$ values for TT purity and TT productivity were both greater than 0.83, while those for FG purity and FG productivity exceeded 0.93. This not only demonstrates that the selected process parameters and material attributes can explain most of the data variability but also indicates, to some extent, the reliability of ChromR's automated experimental execution. Furthermore, ChromR analyzed the developed models and provided reasonable improvement strategies for different performance indicators. Regarding enhancing the purities of TT and FG, ChromR recommended increasing wash time and wash flow. This suggestion is rational: appropriately increasing the wash volume during the wash stage effectively removes highly polar impurities, thereby improving the purity of the less polar target compounds (TT and FG).

3.3.2 Multi-objective optimization

After exploring the process parameter space and establishing predictive models, we proceeded to enable ChromR to perform multi-objective optimization based on the built models. In this work, two batches of feed solutions—250401 and 250409—were selected for validation of multi-objective optimization. Similarly, we described these multi-objective optimization problems using simple prompts and assigned them to ChromR for autonomous optimization. The prompt for batch 250401 is shown in Fig. S18.

Within less than 1 min, ChromR generated five distinct Pareto-optimal solutions for each multi-objective optimization problem, as detailed in Table S3. From these, we selected one optimal solution from each batch for experimental validation by ChromR—solution No. 1 for 250401 and solution No. 3 for 250409. Additionally, we requested ChromR to compute the design space for process and material parameters. The design space was defined such that TT purity must exceed 6%, FG purity must exceed 24%, TT productivity must exceed 50 mg/h, and FG productivity must exceed 200 mg/h. For batch 231201, we also selected a point outside the design space for experimental validation by ChromR. Notably, both selected Pareto-optimal solutions lie within the design space. The process parameters for the validation points are listed in Table S4, the experimental results are summarized in Table 2, and the design space validation results are presented in Fig. 6.

**Table 2.** Experimental validation results.

| Feed solution batch No. | $Y_1$ | | $Y_2$ | | $Y_3$ | | $Y_4$ | | Position in the design space |
|---|---|---|---|---|---|---|---|---|---|
| | Predicted value | Experimental value | Predicted value | Experimental value | Predicted value | Experimental value | Predicted value | Experimental value | |
| 250401 | 6.80 | 7.81 | 96.5 | 98.0 | 29.4 | 32.8 | 380 | 412 | Inside |
| 250409 | 8.52 | 8.28 | 88.6 | 89.2 | 37.5 | 33.8 | 355 | 364 | Inside |
| 231201 | 2.47 | 1.98 | 38.0 | 43.4 | 8.54 | 7.99 | 157 | 175 | Outside |

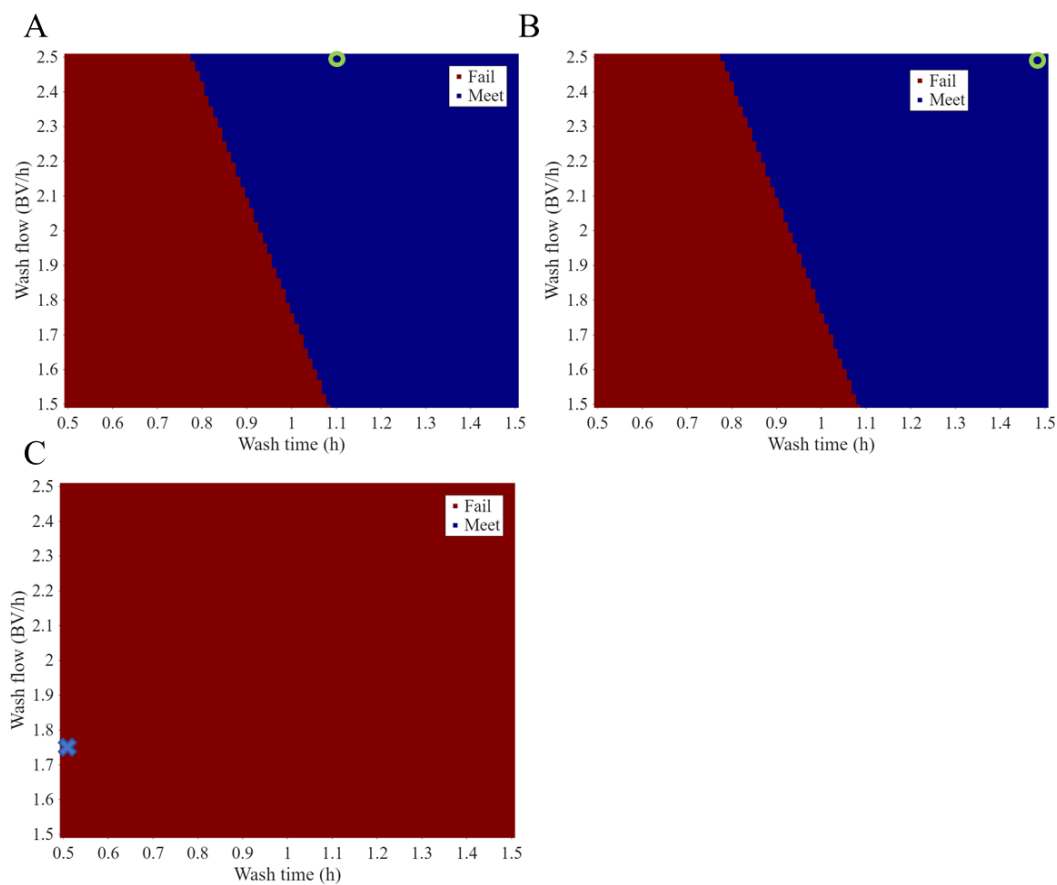

**Fig. 6.** Design space validation results. (A) Batch 250401, $X_1$ = 1.50 BV/h, $X_2$ = 2.00 h, $X_5$ = 3.50 BV/h, $X_6$ = 0.86 h. (B) Batch 250409, $X_1$ = 1.50 BV/h, $X_2$ = 2.00 h, $X_5$ = 3.50 BV/h, $X_6$ = 0.81 h. (C) Batch 231201, $X_1$ = 0.75 BV/h, $X_2$ = 1.00 h, $X_5$ = 3.25 BV/h, $X_6$ = 0.50 h. Symbols × represent points outside the design space, and symbols ○ represent points inside the design space.

Both the Pareto-optimal solutions and the design space were successfully validated. Under the optimized process conditions, both batches 250401 and 250409 surpassed the predefined purity and productivity targets, confirming the reliability of the design space and demonstrating ChromR's excellent performance in chromatographic process optimization.

4. Discussion

4.1 Innovation and value of ChromR

ChromR represents the first application of LLMs to overcome key bottlenecks in traditional chromatographic process development—namely, reliance on expert experience, labor intensity, and long development cycles. During training, ChromLLM fully leveraged extensive, literature-validated prior knowledge, enabling it to autonomously recommend feasible initial process parameters grounded in such knowledge. Moreover, ChromLLM, enhanced with RAG and a literature retrieval module forming Agent A, further improves the reasonableness and advancement of parameter recommendations, significantly reducing dependence on expert judgment and the number of experimental trials required. Secondly, the integration of a multi-agent system with an automated chromatography device enables ChromR to autonomously execute experiments and collect data—greatly reducing manual labor and shortening the overall development timeline.

Conventional chromatographic process development typically involves literature review (1 week), resin screening (2 weeks), OFAT (1 week), DOE (2 weeks), and process optimization and verification (1 week), totalling approximately 7 weeks[61]. In contrast, the ChromR-based development paradigm eliminates the need for the first three steps through Agent A's initial process design and accelerates DOE and process validation via automated experimentation. Taking EGBL as an example, we achieved process parameters meeting predefined objectives within just one week, demonstrating a substantial increase in efficiency.

4.2 Limits

However, there remain aspects of ChromR proposed in this work that can be further improved. Specifically: (1) the absence of an online mixer prevents continuous gradient variation in the mobile phase composition; (2) automatic sample preparation and component analysis have not yet been achieved; (3) ChromR is currently limited to polynomial statistical modelling and cannot establish mechanistic or hybrid models.

5. Conclusion

LLMs are integrated with laboratory automation technologies to construct ChromR, an LLM-driven platform for chromatographic process optimization. ChromR serves as a general-purpose platform for chromatographic process development, effectively addressing long-standing challenges in traditional chromatography development, such as reliance on expert experience, prolonged development cycles, and labor-intensive operations. By leveraging ChromLLM to deeply extract prior knowledge from scientific literature and regulatory guidelines, ChromR enables intelligent generation of initial chromatographic parameters. Furthermore, the integration of a multi-agent system with an automated chromatographic experimental apparatus significantly reduces human labor while substantially improving experimental efficiency and data quality. In the case of EGBL chromatographic process optimization, ChromR successfully considered both fraction purity and productivity requirements, reducing the process development cycle to 1/7 of that required by the conventional paradigm. A reusable technical pathway and engineering practice reference are provided in this work for the intelligent transformation of chromatographic process optimization paradigms.

**CRediT authorship contribution statement**

Zhilong Tang: Conceptualization, Data curation, Formal analysis, Investigation, Methodology, Validation, Visualization, Writing – original draft, Writing – review and editing. ShaoHua Wu: Data curation. Xinyan Zhao: Writing – review and editing. Yu Wang: Data curation. Xingchu Gong: Conceptualization, Funding acquisition, Project administration, Resources, Supervision, Writing – review and editing.

**Declaration of Competing Interest**

The authors declare that they have no known competing financial interests or personal relationships that could have appeared to influence the work reported in this paper.

**Acknowledgements**

This research was funded by the National Key Research and Development Program (2024YFC3506900).

The authors are grateful to Prof. Cheng Yiyu for his useful suggestions. The authors are grateful to Li Tiantian, Feng Yuxuan, and Li Jinhao for their contributions to data annotation.

**Appendix A. Supplementary data**

The following are the Supplementary data to this article:

Supplementary Data.docx

**Data availability**

All codes utilized in this study are openly accessible on GitHub at https://github.com/tzl1125/ChromR.

Supporting information (SI) for:

A Chromatographic Process Design and Optimization Platform Powered by Large Language Models: A Case Application on Extract of *Ginkgo Biloba* Leaf

Zhilong Tang[1], Shaohua Wu[1], Xinyan Zhao[2], Yu Wang[1,3], Xingchu Gong[1,2,3,4,*]

1. Pharmaceutical Informatics Institute, College of Pharmaceutical Sciences, Zhejiang University, Hangzhou 310058, China

2. Institute of Traditional Chinese Medicine, Tianjin University of Traditional Chinese Medicine, Tianjin, 301617, China

3. Jinhua Institute of Zhejiang University, Jinhua 321016, China

4. National Key Laboratory of Chinese Medicine Modernization, Zhejiang University, Hangzhou 310058, China

[*] Correspondence: gongxingchu@zju.edu.cn; Tel.: +86-571-88208427

1. Post-pretraining

The loss curve for post-pretraining of Qwen2.5-14B is shown in Fig. S1.

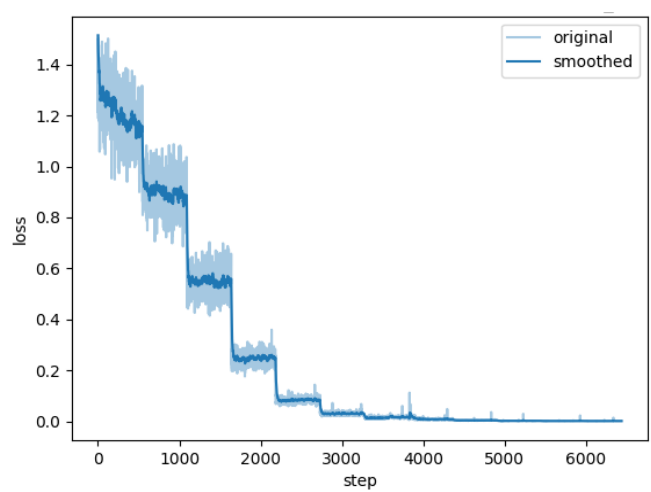

Fig. S1. Loss curve for post-pretraining.

2. Supervised fine-tuning

The loss curve for supervised fine-tuning of the large language model (LLM) after post-pretraining is shown in Fig. S2.

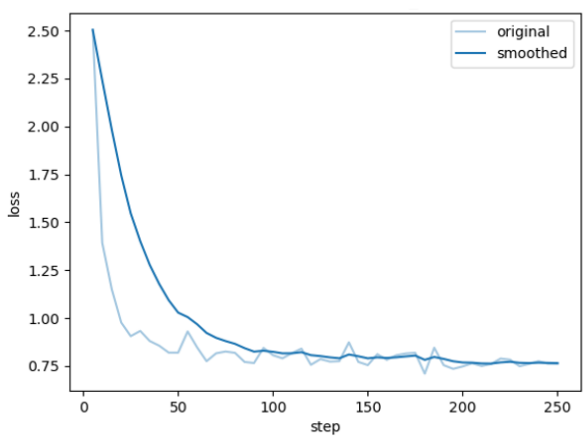

Fig. S2. Loss curve for supervised fine-tuning.

3. Automated chromatography experimental device

A photograph of the automated chromatography experimental device is shown in Fig. S3.

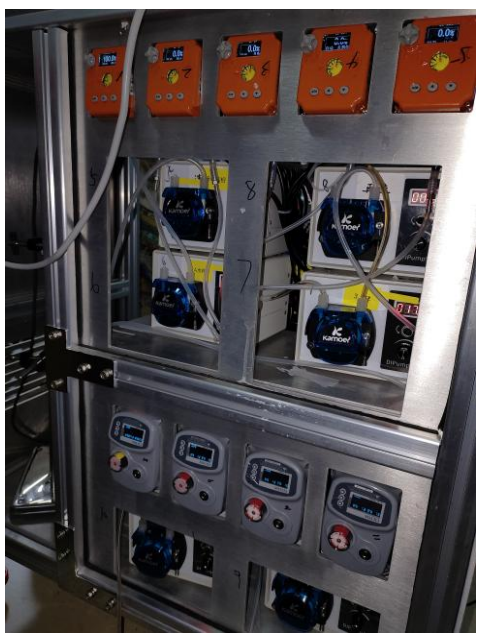

Fig. S3. Automated chromatography experimental device.

4. Web application

A screenshot of the front-end of the ChromR web application is shown in Fig. S4. The front-end includes a dialog box for interacting with the multi-agent system, task execution bar, experiment logs, experiment records, hardware control system, online sensor data, and online spectral data.

Fig. S4. Screenshot of the front-end of the ChromR web application.

5. Materials and methods

5.1 Reagents and materials

Quercetin (lot: 231227, ≥99 %), kaempferol (lot: 240225, ≥98 %), isorhamnetin (lot: 240602, ≥99 %), ginkgolide A (lot: 200616, ≥99 %), ginkgolide B (lot: 191224, ≥99 %), ginkgolide C (lot: 200729, ≥99 %) and bilobalide (lot: 191221, ≥98 %) were all purchased from Shanghai Winherb Medical Science Co., Ltd (Shanghai, China). Formic acid (≥99 %, chromatographic grade, ROE Scientific Inc., Newark, USA), methanol (chromatographic grade, Anhui Tedia High Purity Solvents Co., Ltd., Anhui, China), hydrochloric acid (36–38%, analytical grade, Sinopharm Chemical Reagents Co., Ltd., Shanghai, China) were used. Experimental water was prepared by an ultrapure water system (Milli-Q, Millipore, America). *Ginkgo biloba* leaf alcohol extract concentrate were produced by Tonghua Guhong Pharmaceutical Co., Ltd. (Jilin, China).

5.2 Preparation of feed solution

The *Ginkgo biloba* leaf ethanol-extracted aqueous precipitate solution, which serves as the feed solution, was prepared as follows. The concentrated alcohol extract of *Ginkgo biloba* leaf was mixed with water at a ratio of 1:4 (v/v) and stirred for 30 minutes. It was then cold precipitated at 4°C for 12 hours, followed by filtration. 13 batches of the concentrated alcohol extract of *Ginkgo biloba* leaf were prepared into corresponding feed solutions using this method. The material properties were characterized by the concentration and purity of flavonoid glycosides (FG) and terpene trilactones (TT). The batch numbers and material properties are listed in Table S1.

Table S1. Batch number and material properties of the feed solution.

| Batch No. | TT concentration (mg/mL) | TT purity (%) | FG concentration (mg/mL) | FG purity (%) |
| --- | --- | --- | --- | --- |
| 250401 | 0.583 | 1.01 | 1.85 | 3.21 |
| 250402 | 0.522 | 0.926 | 1.80 | 3.19 |
| 250403 | 0.559 | 0.914 | 1.88 | 3.09 |
| 250404 | 0.495 | 0.966 | 1.72 | 3.36 |
| 250405 | 0.530 | 0.949 | 2.01 | 3.60 |
| 250406 | 0.602 | 1.10 | 1.96 | 3.59 |
| 250407 | 0.570 | 0.948 | 1.85 | 3.07 |
| 250408 | 0.520 | 0.920 | 1.72 | 3.04 |
| 250409 | 0.568 | 0.894 | 1.80 | 2.83 |
| 250501 | 0.502 | 0.874 | 1.69 | 2.94 |
| 231102 | 0.484 | 0.944 | 1.57 | 3.06 |
| 231201 | 0.591 | 1.14 | 1.81 | 3.50 |
| 231202 | 0.641 | 1.16 | 1.96 | 3.55 |

5.3 Sample solution preparation

Preparation method for the test sample solution: 1 mL of *Ginkgo biloba* leaf concentrated solution was taken, followed by the addition of 1 mL of 25% hydrochloric acid and 4 mL of methanol. The mixture was heated under reflux in an 80 °C water bath for 30 minutes, then cooled to room temperature. Methanol was added to bring the total volume to 10 mL. After filtration through a 0.22 μm filter membrane, the subsequent filtrate was used as the test sample solution.

Preparation method for the mixed reference substance solution: Appropriate amounts of quercetin, kaempferol, isorhamnetin, ginkgolide A, ginkgolide B, ginkgolide C, and bilobalide reference standards were accurately weighed respectively and placed in a 10 mL volumetric flask. After dissolution and volume fixation with methanol solution, the flask was placed in an ultrasonic cleaner for 5 minutes of ultrasonication. A precise volume of the reference standard stock solution was pipetted into volumetric flasks and diluted with methanol to prepare a series of reference standard solutions with different concentrations. After filtration through a 0.22 μm filter membrane, the subsequent filtrate was used as the reference standard stock solution.

5.4 Analysis conditions

Samples were analyzed according to the LC/MS method [1]. Mass spectrometry conditions: The single quadrupole mass spectrometer detector is an Agilent InfinitylabLC/MSDiQ (Agilent Technologies, USA), equipped with an ESI ion source, data acquisition in negative ion mode, monitoring m/z 301 for quercetin, m/z 315 for isorhamnetin, m/z 285 for kaempferol, m/z 325 for bilobalide, m/z 423 for ginkgolide B, m/z 439 for ginkgolide C, all monitored as $[M-H]^-$. For ginkgolide A, monitor m/z 453 as $[M-H+COOH]^-$. The nebulizer pressure is 50 psi, the capillary voltage is 2000 V, and the drying gas temperature is 325°C. The fragmentation voltage for bilobalide and ginkgolide A is 90 eV, for ginkgolide B and ginkgolide C it is 100 eV, and for quercetin, kaempferol, and isorhamnetin, it is 140 eV.

Chromatographic conditions: Analysis was performed using an Agilent 1290 HPLC system (Agilent Technologies, USA) with an Agilent InfinityLab Poroshell 120 column (4.6 × 100 mm, 2.7 μm, Agilent Technologies, USA). The mobile phase consisted of Phase A (0.1% formic acid in water) and Phase B (0.1% formic acid in methanol) at a flow rate of 0.66 mL/min, with the column temperature maintained at 30 °C. Gradient elution was employed with the following proportions: 0–15 min, 0–35% B; 15–30 min, 35%–80% B; 30–31 min, 80%–95% B; 31–41 min, 95%–95% B. Chromatograms of the mixed standard and eluent are shown in Fig. S5.

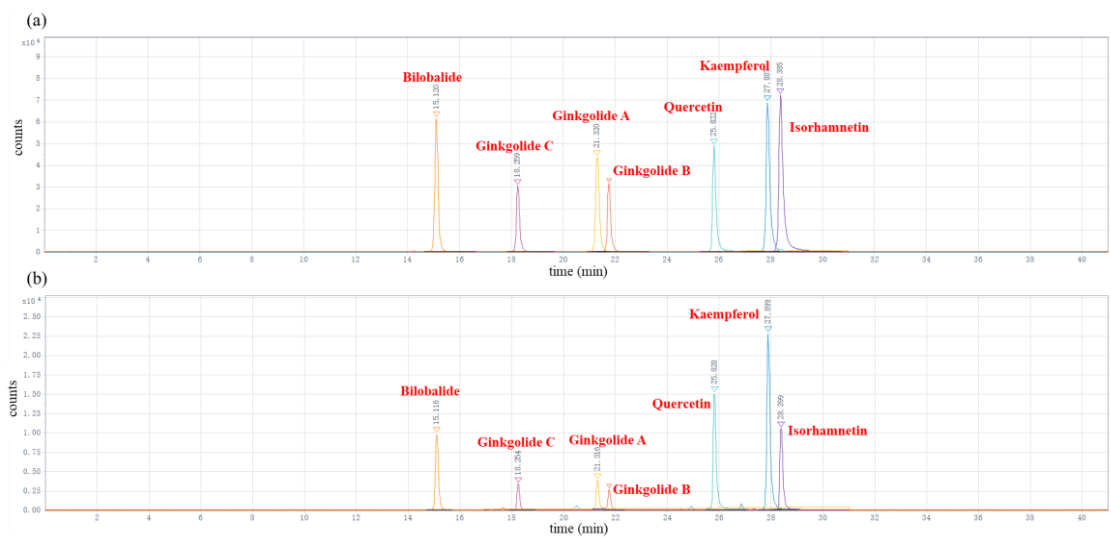

Fig. S5. Chromatograms of the mixed standard (a) and eluent (b).

5.5 Total solid determination

The method for total solid determination refers to the "Pharmacopoeia of the People's Republic of China"[2]. 4 mL of the test solution was accurately added to a pre-weighed weighing bottle that had reached constant weight. The weighing bottle was then placed back into a blast drying oven and dried at 105 °C for at least 3 hours. After drying, it was transferred to a desiccator and allowed to stand for 30 minutes before being taken out. The total mass of the weighing bottle was weighed, and the bottle was returned to the drying oven for further drying. This process was repeated until the difference between two consecutive total masses was within 0.3 mg, and the final mass was recorded. The mass of total solids was calculated as the final mass minus the mass of the empty weighing bottle.

6. Optimization process of the chromatographic process for extract of *Ginkgo biloba* leaf

The following is the optimization process of the chromatographic process, including the dialogue between researchers and ChromR, and the data analysis results provided by ChromR.

## Conversation

Researcher:

In the column chromatography process for preparing Ginkgo biloba leaf extract, if—after sample loading and adsorption is complete—only one isocratic wash followed by one isocratic elution is performed, and the eluate is collected as the target product, then:

1. What type of macroporous resin should be selected to maximize the content of total flavonol glycosides and terpene lactones in the product?
2. What ethanol concentrations are recommended for the wash solution and the elution solution, respectively?

ChromR:

Agent A Execution Result:
In the column chromatography process for preparing Ginkgo biloba leaf extract, if only one isocratic wash and one isocratic elution are performed, and the eluate is collected as the target product, it is essential to holistically consider the type of macroporous resin and the elution conditions in order to maximize the content of **total flavonol glycosides** and **terpene lactones** (primarily including ginkgolides A, B, C, and bilobalide) in the final product.

---

### ☑ Recommended Protocol:

#### **1. Macroporous Adsorption Resin Type: AB-8**
- **Rationale**:
  - AB-8 is a nonpolar to weakly polar polystyrene-divinylbenzene copolymer macroporous resin, offering excellent adsorption capacity and selectivity for enriching and separating moderately polar compounds such as flavonoids and terpene lactones.
  - According to expert chromatographic models and practical process studies, AB-8 resin demonstrates high adsorption capacity and desorption efficiency for total flavonol glycosides and terpene lactones in Ginkgo biloba leaf extracts.
  - Although the Chinese Pharmacopoeia does not specify a particular resin type, it recommends the use of "macroporous adsorption resins" with sequential elution using water followed by ethanol gradients of varying concentrations. AB-8 is one of the standard resins widely adopted in industrial Ginkgo extraction processes.

  > Additionally, literature [2] reports that XDA-1 resin also exhibits excellent performance under pH 2.0 conditions; however, AB-8 is generally preferred in routine production due to its lower cost, superior stability, and well-established application history.
---

#### **2. Ethanol Concentration in Wash Solution: 20% (v/v)**
- **Purpose**: To remove water-soluble impurities (e.g., sugars, certain organic acids, tannins, etc.) while retaining the target compounds on the resin.
- **Mechanism**: A 20% ethanol solution effectively elutes weakly adsorbed impurities, whereas total flavonol glycosides and terpene lactones remain largely retained on the resin at this concentration and are not significantly desorbed.
- **Basis**: Experimental studies indicate that ethanol concentrations below 30% have minimal impact on the target components while effectively enhancing the purity of the subsequent eluate [2].

---

*Continuation*

Fig. S6. The process parameters initially recommended by ChromR.

**Conversation**

--- - - - - - - - - - - - - - Continuation - - - - - - - - - - - - - - - -

ChromR:

#### **3. Ethanol Concentration in Elution Solution: 75% (v/v)**
- **Purpose**: To maximize the elution of total flavonol glycosides and terpene lactones.
- **Mechanism**:
  - Total flavonol glycosides predominantly exist in glycosidic forms with moderate polarity; 75% ethanol provides an optimal polarity environment for efficient desorption.
  - Terpene lactones (e.g., ginkgolides A, B, C, and bilobalide), although relatively less polar, also achieve high recovery rates in 75% ethanol.
- **Experimental Evidence**:
  - Multiple studies have shown that 70–80% ethanol is the optimal range for eluting Ginkgo flavonoids and lactones, with 75% representing the best compromise to balance high yields of both compound classes [2].
  - If the ethanol concentration is too high (e.g., ≥90%), solubility of certain flavonoid glycosides may decrease, and more lipophilic impurities could co-elute; if too low, incomplete elution occurs.

---

### ☑ Summary of Final Recommended Parameters:

| Parameter | Recommended Value | Notes |
|---|---|---|
| **Macroporous Resin Type** | **AB-8** | Nonpolar/weakly polar; suitable for enrichment of flavonoids and lactones |
| **Ethanol Concentration in Wash Solution** | **20%** | Removes impurities while retaining target compounds on the resin |
| **Ethanol Concentration in Elution Solution** | **75%** | Enables efficient co-elution of total flavonol glycosides and terpene lactones |

---

### 🔬 Additional Notes:
- If selective enrichment of a specific compound class (e.g., prioritizing terpene lactones or flavonoids) is desired, consider fractionated collection or pH optimization (e.g., adjusting the sample loading solution to pH 2.0–3.0 to enhance adsorption) [2].
- Under the process constraint of "only one isocratic wash followed by one isocratic elution," the above parameters achieve maximal enrichment of target components while maintaining operational simplicity.

---

### 📑 References

[1] Lejri, I., Vukalović, I., Grimm, A., & Eckert, A. (2025). Proanthocyanidins from Ginkgo extract EGb 761® improve bioenergetics and stimulate neurite outgrowth in vitro. *Frontiers in Pharmacology*, 16, 1495997. https://doi.org/10.3389/fphar.2025.1495997

[2] Zhang, S., Gong, X., & Qu, H. (2024). An effective and comprehensive optimization strategy for preparing Ginkgo biloba leaf extract enriched in shikimic acid by macroporous resin column chromatography. *Phytochemical Analysis*, 35(5), 1234–1245. https://doi.org/10.1002/pca.3375

> Note: Other literature sources were not directly relevant to resin selection or elution concentration optimization and are therefore not cited.

Fig. S7. Continuation of Fig. S6.

# Conversation

Researcher:

In the column chromatography purification process for Ginkgo biloba leaf extract, AB-8 macroporous resin is used as the adsorbent, and the sample loading solution is prepared at a concentration of 0.366 g of herbal material per mL. After sample loading and adsorption, an isocratic wash with 20% (v/v) ethanol followed by an isocratic elution with 75% (v/v) ethanol is performed, with the eluate collected as the target product.

To minimize processing time per cycle and maximize the contents of total flavonol glycosides and terpene lactones in the final product—while simultaneously reducing material waste—please address the following operational parameters:

1. **What should the bed height-to-diameter ratio (H/D) be?**
2. **What are the recommended volumes (in column volumes, BV) for sample loading, washing, and elution, respectively?**
3. **What flow rates (e.g., in BV/h or mL/min) are recommended for the loading, washing, and elution stages, respectively?**

---

ChromR:

**Agent A Execution Result:**

In the column chromatography purification process for *Ginkgo biloba* leaf extract, **AB-8 macroporous resin** is employed as the adsorbent. The objective is to **minimize material waste while achieving the shortest possible single-cycle processing time and maximizing the contents of total flavonol glycosides and terpene lactones in the final product**. Based on the preparation requirements specified in the *Chinese Pharmacopoeia* and optimization recommendations provided by a domain-specific large model for chromatography, the following key process parameters are recommended:

---

### 1. Bed Height-to-Diameter Ratio (H/D)

**Recommended bed height-to-diameter ratio: 12:1** (e.g., 12 cm bed height / 1 cm inner column diameter).

This H/D ratio enhances column efficiency and separation selectivity, reduces axial dispersion while ensuring adequate mass transfer, and improves enrichment of flavonol glycosides and terpene lactones. It balances column packing stability, flow uniformity, and elution resolution—making it suitable for industrial scale-up.

> **Note**: An excessively low H/D ratio leads to incomplete separation, while an overly high ratio may cause excessive pressure drop and difficulties in flow control.

---

— — — — — — — — — — — — — *Continuation* — — — — — — — — — — — — —

Fig. S8. Other process parameters recommended by ChromR.

ChromR:

### 2. Loading, Washing, and Elution Volumes (expressed in Column Volumes, BV)

| Step        | Recommended Volume      |
|-------------|-------------------------|
| **Loading** | **1.5 BV**              |
| **Washing** | **2.0 BV** (20% ethanol)|
| **Elution** | **3.0 BV** (75% ethanol)|

- **Loading at 1.5 BV**: At the given sample concentration (0.366 g herbal material/mL), this volume fully utilizes the resin's adsorption capacity without overloading, thereby preventing breakthrough of target compounds.
- **Washing with 2.0 BV of 20% ethanol**: Effectively removes water-soluble impurities (e.g., sugars, certain phenolic acids) while retaining flavonol glycosides and terpene lactones on the resin.
- **Elution with 3.0 BV of 75% ethanol**: Ensures complete desorption of strongly bound flavonol glycosides and terpene lactones, maximizing recovery while producing a concentrated eluate that facilitates downstream concentration.

---

### 3. Recommended Flow Rates (in BV/h)

| Stage       | Recommended Flow Rate |
|-------------|-----------------------|
| **Loading** | **1 BV/h**            |
| **Washing** | **2 BV/h**            |
| **Elution** | **3 BV/h**            |

- **Loading at 1 BV/h**: A slower flow rate allows sufficient diffusion of target compounds into the resin pores, improving adsorption efficiency and utilization of binding capacity.
- **Gradually increased flow rates during washing and elution**: Since washing primarily removes loosely bound impurities and 75% ethanol enables rapid mass transfer during elution, moderately higher flow rates do not compromise recovery but significantly reduce overall cycle time and improve throughput.

---

### 4. Expected Performance

Under the above conditions, model predictions indicate:

- **Total flavonol glycosides content: 24.79%**
- **Terpene lactones content: 6.42%**

These values meet the quality specifications outlined in the *Chinese Pharmacopoeia* for *Ginkgo biloba* leaf extract (total flavonol glycosides ≥24%; terpene lactones ≥6%), achieving an **optimal balance among product quality, production efficiency, and material utilization**.

---

— — — — — — — — — — — — — Continuation — — — — — — — — — — — — —

Fig. S9. Continuation of Fig. S8.

**Conversation**

--- *Continuation* ---

ChromR:

### 5. Conclusion

To enable efficient, low-waste, and high-quality purification of *Ginkgo biloba* leaf extract, the following process parameters are recommended:

| Parameter | Recommended Value |
|---|---|
| Bed height-to-diameter ratio | 12:1 (H/D) |
| Loading volume | 1.5 BV |
| Washing volume | 2.0 BV (20% ethanol) |
| Elution volume | 3.0 BV (75% ethanol) |
| Loading flow rate | 1 BV/h |
| Washing flow rate | 2 BV/h |
| Elution flow rate | 3 BV/h |

This protocol is designed based on the adsorption characteristics of macroporous resins and polarity differences among constituents. It aligns with the preparation principles of the *Chinese Pharmacopoeia* [1] and has been validated through model-based optimization, making it suitable for standardized production at both laboratory and pilot scales.

---

### References

[1] Chinese Pharmacopoeia Commission. *Pharmacopoeia of the People's Republic of China (2020 Edition, Volume I)*: Vegetable Oils and Extracts – *Ginkgo biloba* Leaf Extract [S]. Beijing: China Medical & Pharmaceutical Press, 2020.

Fig. S10. Continuation of Fig. S9.

# Conversation

Researcher:

Run a DSD experimental design. The loading solution batches include ten lots: 231102, 231202, 250402, 250403, 250404, 250405, 250406, 250407, 250408, and 250501. The loading flow rate is set at 0.5–1.5 BV/h, with a loading time of 1–2 h. The wash flow rate is 1.5–2.5 BV/h, and the wash time is 0.5–1.5 h. The elution flow rate is 2.5–3.5 BV/h, and the elution time is 0.5–1.5 h. Two dummy factors are included, and three center point experiments are added.

ChromR:

Agent B Execution Result:

| No. | Loading Flow Rate | Loading Time | Wash Flow Rate | Wash Time | Elution Flow Rate | Elution Time | Batch Number |
|---|---|---|---|---|---|---|---|
| 1 | 1.0 | 2.0 | 2.5 | 1.5 | 3.5 | 1.5 | 250408 |
| 2 | 1.0 | 1.0 | 1.5 | 0.5 | 2.5 | 0.5 | 250403 |
| 3 | 1.5 | 1.5 | 1.5 | 0.5 | 3.5 | 0.5 | 231102 |
| 4 | 0.5 | 1.5 | 2.5 | 1.5 | 2.5 | 1.5 | 231202 |
| 5 | 1.5 | 2.0 | 2.0 | 0.5 | 2.5 | 1.5 | 250405 |
| 6 | 0.5 | 1.0 | 2.0 | 1.5 | 3.5 | 0.5 | 250405 |
| 7 | 1.5 | 2.0 | 2.5 | 1.0 | 2.5 | 0.5 | 250404 |
| 8 | 0.5 | 1.0 | 1.5 | 1.0 | 3.5 | 1.5 | 250402 |
| 9 | 1.5 | 1.0 | 2.5 | 1.5 | 3.0 | 0.5 | 250408 |
| 10 | 0.5 | 2.0 | 1.5 | 0.5 | 3.0 | 1.5 | 250406 |
| 11 | 1.5 | 2.0 | 1.5 | 1.5 | 3.5 | 1.0 | 250402 |
| 12 | 0.5 | 1.0 | 2.5 | 0.5 | 2.5 | 1.0 | 250407 |
| 13 | 1.5 | 1.0 | 2.5 | 0.5 | 3.5 | 1.5 | 250406 |
| 14 | 0.5 | 2.0 | 1.5 | 1.5 | 2.5 | 0.5 | 231202 |
| 15 | 1.5 | 1.0 | 1.5 | 1.5 | 2.5 | 1.5 | 250501 |
| 16 | 0.5 | 2.0 | 2.5 | 0.5 | 3.5 | 0.5 | 250403 |
| 17 | 1.0 | 1.5 | 2.0 | 1.0 | 3.0 | 1.0 | 250407 |
| 18 | 1.0 | 1.5 | 2.0 | 1.0 | 3.0 | 1.0 | 250501 |
| 19 | 1.0 | 1.5 | 2.0 | 1.0 | 3.0 | 1.0 | 250404 |
| 20 | 1.0 | 1.5 | 2.0 | 1.0 | 3.0 | 1.0 | 231102 |

Fig. S11. Experimental design table provided by ChromR.

**Researcher:**

Create an experiment, column height 45.6 cm, column inner diameter 3 cm, liquid level height 3.0 cm, bed height 36 cm, resin AB-8, loading solution batch number 250401, wash solution 20% aqueous ethanol, elution solution 75% aqueous ethanol, regeneration solution 95% aqueous ethanol, equilibration solution pure water. Loading flow rate 1.5 BV/h, loading time 2.0 h, wash flow rate 2.5 BV/h, wash time 1.095 h, elution flow rate 3.5 BV/h, elution time 0.86 h. Regeneration and equilibration flow rates both 3 BV/h, fraction selected "F1".

**ChromR:**

Agent C execution result:
Experiment created successfully, ID is 33.

Fig. S12. Creating an experiment through prompts in ChromR.

**Researcher:**

Run experiment id 33.

**ChromR:**

Agent C execution result:
Start executing experiment ID 33.

Fig. S13. Running a specified experiment through prompts in ChromR.

# Conversation

Researcher:

Create a DSD experimental design table, column height 45.6 cm, column inner diameter 3 cm, liquid level height 3.0 cm, bed height 36 cm, resin AB-8, wash solution 20% aqueous ethanol, elution solution 75% aqueous ethanol, regeneration solution 95% aqueous ethanol, equilibration solution pure water. Regeneration and equilibration flow rates both 3 BV/h, fraction selected "F1".

The experimental design table is as follows:

| No. | Loading Flow Rate | Loading Time | Wash Flow Rate | Wash Time | Elution Flow Rate | Elution Time | Batch Number |
|---|---|---|---|---|---|---|---|
| 1 | 1.0 | 2.0 | 2.5 | 1.5 | 3.5 | 1.5 | 250408 |
| 2 | 1.0 | 1.0 | 1.5 | 0.5 | 2.5 | 0.5 | 250403 |
| 3 | 1.5 | 1.5 | 1.5 | 0.5 | 3.5 | 0.5 | 231102 |
| 4 | 0.5 | 1.5 | 2.5 | 1.5 | 2.5 | 1.5 | 231202 |
| 5 | 1.5 | 2.0 | 2.0 | 0.5 | 2.5 | 1.5 | 250405 |
| 6 | 0.5 | 1.0 | 2.0 | 1.5 | 3.5 | 0.5 | 250405 |
| 7 | 1.5 | 2.0 | 2.5 | 1.0 | 2.5 | 0.5 | 250404 |
| 8 | 0.5 | 1.0 | 1.5 | 1.0 | 3.5 | 1.5 | 250402 |
| 9 | 1.5 | 1.0 | 2.5 | 1.5 | 3.0 | 0.5 | 250408 |
| 10 | 0.5 | 2.0 | 1.5 | 0.5 | 3.0 | 1.5 | 250406 |
| 11 | 1.5 | 2.0 | 1.5 | 1.5 | 3.5 | 1.0 | 250402 |
| 12 | 0.5 | 1.0 | 2.5 | 0.5 | 2.5 | 1.0 | 250407 |
| 13 | 1.5 | 1.0 | 2.5 | 0.5 | 3.5 | 1.5 | 250406 |
| 14 | 0.5 | 2.0 | 1.5 | 1.5 | 2.5 | 0.5 | 231202 |
| 15 | 1.5 | 1.0 | 1.5 | 1.5 | 2.5 | 1.5 | 250501 |
| 16 | 0.5 | 2.0 | 2.5 | 0.5 | 3.5 | 0.5 | 250403 |
| 17 | 1.0 | 1.5 | 2.0 | 1.0 | 3.0 | 1.0 | 250407 |
| 18 | 1.0 | 1.5 | 2.0 | 1.0 | 3.0 | 1.0 | 250501 |
| 19 | 1.0 | 1.5 | 2.0 | 1.0 | 3.0 | 1.0 | 250404 |
| 20 | 1.0 | 1.5 | 2.0 | 1.0 | 3.0 | 1.0 | 231102 |

ChromR:

Agent C execution result:
Experimental design table created successfully, experimental design table ID: 1, experiment IDs: [13,14,15,16,17,18,19,20,21,22,23,24,25,26,27,28,29,30,31,32].

Fig. S14. Creating an experimental design table through prompts in ChromR.

## Conversation

Researcher:

Please perform stepwise regression analysis on the experimental results (with a p-value threshold of 0.05 for term selection) and provide predictive models for each response variable. Additionally, please identify the most influential terms in the models, interpret their practical significance, and suggest methods to improve the response variables.

Upload files: Experiment results  Parameter condition  Material condition

ChromR:

Agent D execution result is as follows:

Based on the stepwise regression analysis results, the predictive models for each response variable, the most influential terms, and methods to improve the responses are as follows:

### 1. Lactone Purity
**Model:**
lactone_purity = 3.1167 − 10.7425 × TT_concentration + 0.7026 × sample_flow + 3.9301 × washing_time + 1.8364 × washing_flow
**Coefficient of Determination ($R^2$):** 0.8538

**Key Influential Terms:**
- **washing_time (3.9301):** Has a significant positive effect on lactone purity.
- **washing_flow (1.8364):** Also has a significant positive effect on lactone purity.

**Improvement Strategies:**
- Increasing washing_time and washing_flow can effectively enhance lactone purity.

### 2. Flavonol Glycoside Purity
**Model:**
flavonol_glycoside_purity = 0.4002 − 50.0259 × TT_concentration + 18.7848 × washing_time + 9.5709 × washing_flow + 3.8743 × elution_time + 2.9674 × elution_flow
**Coefficient of Determination ($R^2$):** 0.9574

**Key Influential Terms:**
- **washing_time (18.7848):** Has a very strong positive impact on flavonol glycoside purity.
- **washing_flow (9.5709):** Also significantly improves flavonol glycoside purity.

**Improvement Strategies:**
- Increasing washing_time and washing_flow can effectively improve flavonol glycoside purity. Additionally, moderately increasing elution_time and elution_flow will also be beneficial.

― ― ― ― ― ― ― ― ― ― *Continuation* ― ― ― ― ― ― ― ― ― ―

Fig. S15. Stepwise regression analysis by ChromR.

# Conversation

― ― ― ― ― ― ― ― ― ― ― ― Continuation ― ― ― ― ― ― ― ― ― ― ― ―

ChromR:

### 3. Lactone Productivity
**Model:**
lactone_productivity = 32.8842 + 17.5758 × (sample_flow × elution_flow) − 2.3691 × sample_time − 23.4526 × sample_flow − 8.2481 × elution_flow − 5.7581 × washing_time − 9.4687 × (sample_flow × washing_time) + 20.6692 × (sample_time × sample_flow)
**Coefficient of Determination ($R^2$):** 0.8377

**Key Influential Terms:**
- **(sample_flow × elution_flow) (17.5758):** Positively affects lactone productivity.
- **sample_flow (−23.4526):** Has a significant negative effect on lactone productivity.

**Improvement Strategies:**
- Enhance the interaction between sample_flow and elution_flow (i.e., increase both simultaneously) to boost lactone productivity. However, sample_flow should be carefully controlled to avoid its strong negative impact.

### 4. Flavonol Glycoside Productivity
**Model:**
flavonol_glycoside_productivity = 47.4225 + 49.6461 × (sample_flow × elution_flow) + 13.6240 × sample_time − 18.6213 × sample_flow − 10.5198 × elution_flow − 22.3586 × washing_time − 26.3807 × (sample_flow × washing_time) + 58.9702 × (sample_time × sample_flow)
**Coefficient of Determination ($R^2$):** 0.9322

**Key Influential Terms:**
- **(sample_flow × elution_flow) (49.6461):** Has a significant positive effect on flavonol glycoside productivity.
- **(sample_time × sample_flow) (58.9702):** Also has a strong positive effect on flavonol glycoside productivity.

**Improvement Strategies:**
- Increase both the interaction of sample_flow with elution_flow and the interaction of sample_time with sample_flow to significantly enhance flavonol glycoside productivity.

### Overall Evaluation:
- The models for **lactone purity** and **flavonol glycoside purity** show good fit, especially flavonol glycoside purity with an $R^2$ of 0.9574, indicating the model explains data variation well.
- The models for **lactone productivity** and **flavonol glycoside productivity** include interaction terms, but some coefficients may have relatively high p-values, suggesting these terms might not all be statistically significant. Further validation or model refinement is recommended.
- **washing_time** and **washing_flow** consistently show significant positive effects across multiple responses; thus, increasing these parameters is an effective strategy to improve product quality in practice.

Fig. S16. Continuation of Fig. S11.

ChromR established response surface models for each indicator using a second-order mathematical model, with the quantitative model shown in Equation S1.

$$Y = b_0 + \sum_{i=1}^{n} b_i X_{c,i} + \sum_{i=1}^{n} b_{ii} X_{c,i}^2 + \sum_{i=1}^{n-1} \sum_{j=i+1}^{n} b_{ij} X_{c,i} X_{c,j} + \sum_{k=1}^{m} c_k Z_{c,k} \qquad (S1)$$

where $n$ and $m$ are the numbers of process parameters and material properties, respectively; $X_{c,i}$, $X_{c,j}$ are the process parameters; $Z_{c,k}$ is the material property; $b_0$ is the constant term; $b_i$, $b_{ii}$, $b_{ij}$ and $c_k$ are the partial regression coefficients. ChromR first performed stepwise regression to screen out unimportant terms, followed by partial least squares fitting. The models obtained for each indicator are shown in Table S2. Process parameters are marked as follows: feed flow (BV/h) as $X_1$, feed time (h) as $X_2$, wash flow (BV/h) as $X_3$, wash time (h) as $X_4$, elution flow (BV/h) as $X_5$, and elution time (h) as $X_6$. Material properties are marked as follows: TT concentration (mg/mL) as $Z_1$, TT purity (%) as $Z_2$, FG concentration (mg/mL) as $Z_3$, and FG purity (%) as $Z_4$. Responses are marked as follows: TT purity (%) as $Y_1$, TT productivity (mg/h) as $Y_2$, FG purity (%) as $Y_3$, and FG productivity (mg/h) as $Y_4$.

Table S2. Partial regression coefficients and determination coefficients of each indicator model.

| Terms | $Y_1$ | $Y_2$ | $Y_3$ | $Y_4$ |
| --- | --- | --- | --- | --- |
| Intercept | 3.1167 | 32.8842 | 0.4002 | 47.4225 |
| $Z_1$ | -10.7425 | | -50.0259 | |
| $X_1$ | 0.7026 | -23.4526 | | -18.6213 |
| $X_2$ | | -2.3691 | | 13.6240 |
| $X_3$ | 1.8364 | | 9.5709 | |
| $X_4$ | 3.9301 | -5.7581 | 18.7848 | -22.3586 |
| $X_5$ | | -8.2481 | | -10.5198 |
| $X_6$ | | | 3.8743 | |
| $X_1 \times X_5$ | | 17.5758 | | 49.6461 |

| Terms | $Y_1$ | $Y_2$ | $Y_3$ | $Y_4$ |
| --- | --- | --- | --- | --- |
| $X_1 \times X_4$ | | -9.4687 | | -26.3807 |
| $X_1 \times X_2$ | | 20.6692 | | 58.9702 |
| $R^2$ | 0.8538 | 0.8377 | 0.9574 | 0.9322 |

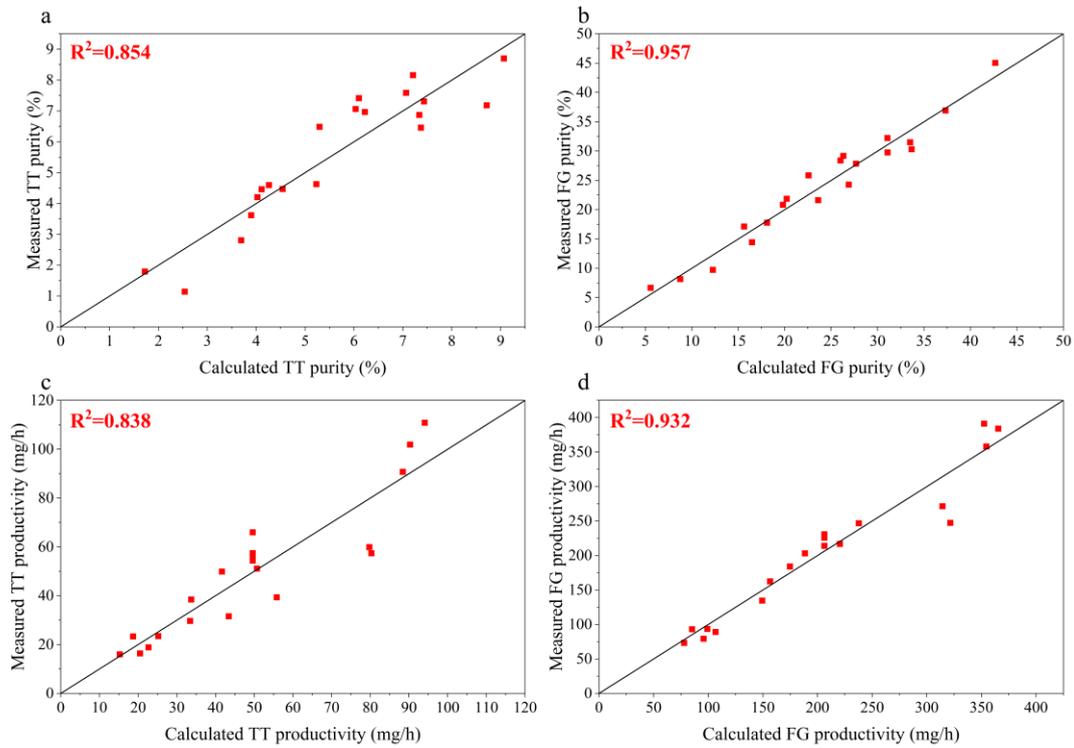

Fig. S17. Measured and calculated values for each indicator. (a) TT purity. (b) FG purity. (c) TT productivity. (d) FG productivity.

## Conversation

Researcher:

Please perform multi-objective optimization
The following are the calculation formulas for each response:
lactone_purity = 3.1167 - 10.7425 × 0.5835 + 0.7026 × sample_flow + 3.9301 × washing_time + 1.8364 × washing_flow
flavonol_glycoside_purity = 0.4002 - 50.0259 × 0.5835 + 18.7848 × washing_time + 9.5709 × washing_flow + 3.8743 × elution_time + 2.9674 × elution_flow
lactone_productivity = 32.8842 + 17.5758 × (sample_flow × elution_flow) - 2.3691 × sample_time - 23.4526 × sample_flow - 8.2481 × elution_flow - 5.7581 × washing_time - 9.4687 × (sample_flow × washing_time) + 20.6692 × (sample_time × sample_flow)
flavonol_glycoside_productivity = 47.4225 + 49.6461 × (sample_flow × elution_flow) + 13.6240 × sample_time - 18.6213 × sample_flow - 10.5198 × elution_flow - 22.3586 × washing_time - 26.3807 × (sample_flow × washing_time) + 58.9702 × (sample_time × sample_flow)
Variable ranges: sample_time[1.00, 2.00], sample_flow[0.5, 1.5], washing_time[0.50, 1.50], washing_flow[1.5, 2.5], elution_time[0.50, 1.50], elution_flow[2.5, 3.5]
Objectives: lactone_purity greater than 6 and maximized, flavonol_glycoside_purity greater than 24 and maximized, lactone_productivity and flavonol_glycoside_productivity both maximized.
Parameters: population size set to 2000, number of iterations 100

ChromR:

Agent D execution result:
Multi-objective optimization task started; details can be viewed in the task status.

Fig. S18. Multi-objective optimization by ChromR.

Table S3. Pareto optimal solutions and predicted indicator values for feed solutions 250401 and 250409.

| Batch No. of feed solution | Solution No. | $X_1$ (BV/h) | $X_2$ (h) | $X_3$ (BV/h) | $X_4$ (h) | $X_5$ (BV/h) | $X_6$ (h) | $Y_1$ (%) | $Y_3$ (%) | $Y_2$ (mg/h) | $Y_4$ (mg/h) |
|---|---|---|---|---|---|---|---|---|---|---|---|
|  | 1 | 1.50 | 2.00 | 2.50 | 1.10 | 3.50 | 0.860 | 6.80 | 29.4 | 96.5 | 380 |
| 250401 | 2 | 1.50 | 2.00 | 2.50 | 1.33 | 3.50 | 0.501 | 7.74 | 32.5 | 91.7 | 365 |
|  | 3 | 1.50 | 2.00 | 2.50 | 1.29 | 3.50 | 1.23 | 7.58 | 34.6 | 92.6 | 367 |

| Batch No. of feed solution | Solution No. | $X_1$ (BV/h) | $X_2$ (h) | $X_3$ (BV/h) | $X_4$ (h) | $X_5$ (BV/h) | $X_6$ (h) | $Y_1$ (%) | $Y_3$ (%) | $Y_2$ (mg/h) | $Y_4$ (mg/h) |
|---|---|---|---|---|---|---|---|---|---|---|---|
| | 4 | 1.50 | 2.00 | 2.50 | 0.892 | 3.50 | 0.515 | 6.00 | 24.3 | 101 | 392 |
| | 5 | 1.50 | 2.00 | 2.50 | 1.40 | 3.50 | 1.33 | 7.99 | 36.9 | 90.4 | 361 |
| | 1 | 1.50 | 2.00 | 2.50 | 1.50 | 3.50 | 1.50 | 8.56 | 40.3 | 88.4 | 355 |
| | 2 | 1.50 | 2.00 | 2.50 | 1.31 | 3.50 | 1.50 | 7.83 | 36.8 | 92.2 | 366 |
| 250409 | 3 | 1.50 | 2.00 | 2.50 | 1.49 | 3.50 | 0.815 | 8.52 | 37.5 | 88.6 | 355 |
| | 4 | 1.50 | 2.00 | 2.50 | 1.45 | 3.50 | 0.501 | 8.38 | 35.6 | 89.4 | 357 |
| | 5 | 1.50 | 2.00 | 2.50 | 1.10 | 3.50 | 1.50 | 6.99 | 32.8 | 96.4 | 379 |

Table S4. Process parameters for the validation experiments.

| Batch No. of feed solution | $X_1$ (BV/h) | $X_2$ (h) | $X_3$ (BV/h) | $X_4$ (h) | $X_5$ (BV/h) | $X_6$ (h) | Position in the design space |
|---|---|---|---|---|---|---|---|
| 250401 | 1.5 | 2 | 2.5 | 1.095 | 3.5 | 0.86 | Inside |
| 250409 | 1.5 | 2 | 2.5 | 1.49 | 3.5 | 0.81 | Inside |
| 231201 | 0.75 | 1 | 1.75 | 0.5 | 3.25 | 0.5 | Outside |